# Automatic Generation of C-code or PLD Circuits under SFC Graphical Environment

Carla Ferreira, Sérgio Monteiro, João Monteiro
Department of Industrial Electronics
University of Minho
4800 Guimarães, Portugal
e-mails: {carla.ferreira, sergio.monteiro, joao.monteiro}@dei.uminho.pt

*Abstract* – This paper proposes a framework for automatic development of control systems from a high level specification based in Grafcet formalism. Grafcet, or Sequential Function Charts (SFC), is a special class of Petri Nets and is becoming the standard representation for sequential control systems. The proposed framework accepts a graphical (through ISaGRAPH) or textual behavioural specification of the control system to be implemented. It follows the usual procedure in software specification: the first step is to formally validate the initial specification. Then the initial specification is translated through automated processes into an implementation. At the moment there are two possible output languages: C and Palasm [1]. The target processor for the C code language are microcontroller based systems that require extended time constrains and access to external peripherals. The goal of including PLD's is the possibility of automatically design mixed hardware and software systems [2].

## I. INTRODUCTION

Petri Nets have shown to be an adequate tool to specify and model control systems [2, 3, 4, 5, 6]. Beside that, the success of this methodology relies on a well study and established mathematical theory, which enables the use of formal validation methods [7].

This paper is structured as follows: A simple overview of Grafcet and rule-based specification formalism is given in section II. Also, in that section we describe briefly the method to translate a grafcet into conditional rules. The framework is described in section III. In section IV it is discussed the consequences of the inclusion of macrosteps in the specification of control systems. In that section it is also discussed the problem of interrupting the execution of a step and the existence of grafcet libraries.

## II. PRELIMINARIES

The framework presented in this paper uses Grafcet, a restricted Petri Net, as the high level specification tool to model control systems [5]. Its main goal is to automatically design the controller from the high level specification, independently on the technological process of the target system.

*A. Grafcet*

Grafcet is a graphical language to represent sequential function charts [5]. Each grafcet is a graph composed by two types of nodes, steps and transitions (a grafcet contains at least one step and one transition). A step may have two states: active or inactive.

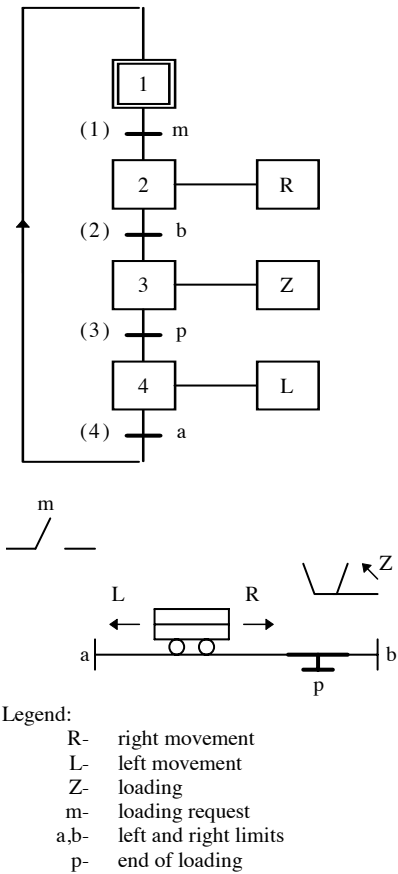

Fig. 1. Process of loading a truck

Legend:
R- right movement
L- left movement
Z- loading
m- loading request
a,b- left and right limits
p- end of loading

Each step has associated actions that will be executed when the step became active. A transition connects two step. Each transition has associated a receptivity. The receptivity is a boolean function of the grafcet input variables.

A transition is enabled when all the steps preceding that transition are active. The receptivity is tested while the transition is enabled.

When the receptivity is true the transition is fired, which means that all the steps preceding that transition are deactivated and all the steps connected after the transition are activated. The example presented in Fig. 1 is used to illustrate these notions.

The complete system may be divided in two units: the processing unit (formed by the processes to be controlled: truck and hopper) and the control unit (the logic controller to be implemented, specified by the grafcet).

The logic controller inputs are m, a, b and p (inputs are associated with transitions either directly or combined within a logical function).

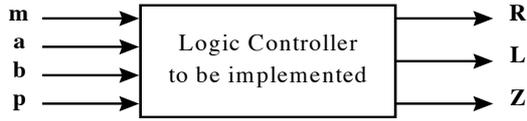

Fig. 2. Interface of the logic controller with inputs and outputs.

The logic controller outputs are R, L and Z which activate the processes to be controlled (outputs are associated with steps). Fig. 2 shows the interface of the logic controller described by the grafcet presented in Fig. 1.

We use the name *Grafcet* to the description language itself, while *grafcet* is used to denote each controller chart.

*B. Rule-based Logic Specification*

Another formalism that can be used to specify discrete control systems is Gentzen Sequent Logic with some elements from temporal logic. Between this logic controller and Grafcet formalism exists a direct correspondence, that is, we can define a bijective function that maps each grafcet into the correspondent logic theory (or reverse operation) [7, 8].

The transformation of a grafcet model into a rule-based logical description follows a simple method: each transition of the grafcet is mapped to a rule, in which the precondition is formed by its input steps and receptivity, and the post-condition is formed by its output steps and output sub-rules. Fig.3 presents the rule-based description of the logic controller specified in Fig. 1.

In this example, the transition t1 has as input step p1, output step p2 and the receptivity function is m. Step p3 has one associated action: Z.

The rule-based description does not describe sequencing, but possible sequences of operation could be derived by ordering the transitions rules.

## III. THE PROPOSED FRAMEWORK

The framework, illustrated in Fig. 4, accepts as input the grafcet or the rule-based specification of the control system.

```
Step                p1, p2, p3, p4
Transition          t1, t2, t3, t4
Input               m, b, p, a
Output              R, Z, L
Marking             p1
Transitions:
    t1: p1*m        |- @p2
    t2: p2*b        |- @p3
    t3: p3*p        |- @p4
    t4: p1*a        |- @p1
Steps:
    p2:             |- R
    p3:             |- Z
    p2:             |- L

where:
 * and, |- yield, @ next    Operators
 p1:, p2:, p3:, p4:         labels of step rules
 t1:, t2:, t3:, t4:         labels of transition rules
```

Fig. 3. Rule-based specification of the logic controller

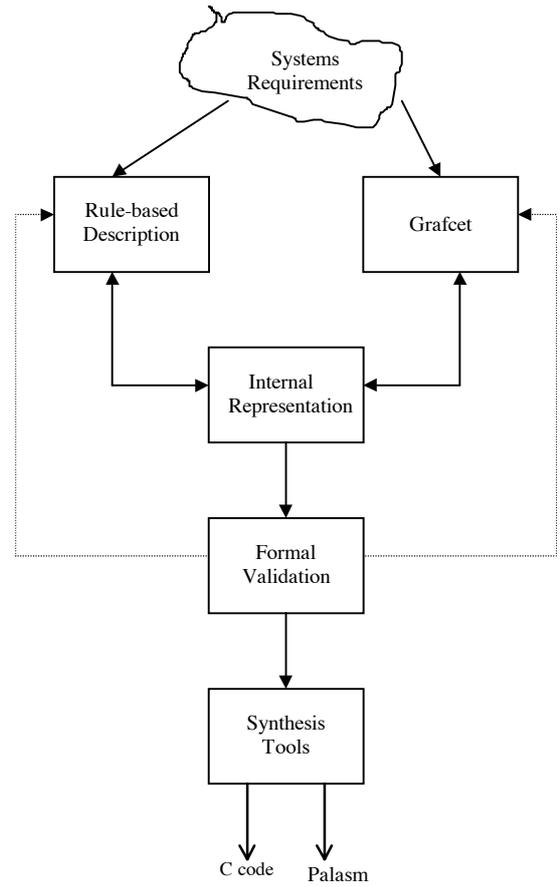

Fig.4. Proposed Framework

With this approach, the user can choose to specify the control system graphically or textually, having the guaranty that these two representations can be used interchangeably.

To edit the grafcet we used ISaGRAPH, a CASE tool that follows the international norm IEC 1131-3 but with no control algorithm analysis. From the graphical representation of the net, ISaGRAPH creates several auxiliary files, which are used to synthesise an internal representation of the net. From this internal representation it can be automatically deduced the rule-based specification of the input grafcet.

Alternatively, any text editor can be used to write the rule-based description of the controller. From this file is extracted its internal representation. Again, through this representation it is possible to obtain automatically the grafcet equivalent to the input description. So, in this framework, it is possible to commute between the graphical and textual representation of the controller.

*A. Formal Validation*

Although Grafcet is a special class of Petri Nets there are differences of behaviour that restrict the application of the usual Petri Nets properties analysis methods to Grafcet.

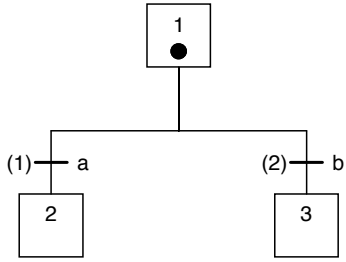

Fig. 5: Conflict between two transitions

In this framework the formal validation consists in verifying if the grafcet is live and conflict free (by definition a grafcet is bounded).

Them main purpose is to translate a grafcet, which was designed based on common sense rules, to another grafcet that is conflict free.

If in the formal verification a conflict is detected in the grafcet we sustain that it doesn't imply an error in the specification. To exemplify this, it will be used the grafcet of Fig. 5. Analysing this grafcet we detect a conflict between transitions (2) and (3): they both depend on a common step and their receptivities may be simultaneously true.

If the receptivity **a**=**b** then the grafcet of Fig. 5 doesn't have any conflict. It states that two parallel activities must initiate at same time. In this case, the grafcet of Fig. 5 can be replaced by the grafcet of Fig. 6, which has the same meaning but a more correct representation. In fact, this new graph copes with the formalism associated with Petri net theory. An experienced programmer would probably design this graph using the parallel and synchronous solution presented in figure 6.

In the case of **a**≠**b** we have the situation where the final result can not be predicted, it depends on the order of occurrence of the events. When step 1 becomes active the final result depends on the order on which receptivity **a** and **b** becomes true. If receptivity **a** becomes true (logic 1) and **b** is false (logic 0), then step 1 is deactivated and step 2 activated. On the opposite case, where **a**=0 and **b**=1, step 3 is activated. Finally, when, and if, both receptivity **a** and **b** became true simultaneously step 2 and step 3 are activated.

If we intended to cope with a strict formalism, this last solution should be avoided. Our experience however points in the opposite direction: commonly, users use a direct translation from a natural language specification that leads to these type of conflicts.

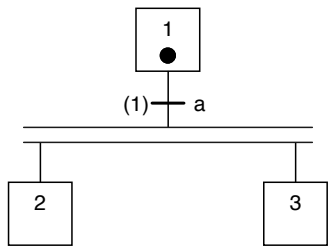

Fig. 6: Grafcet for parallel activities.

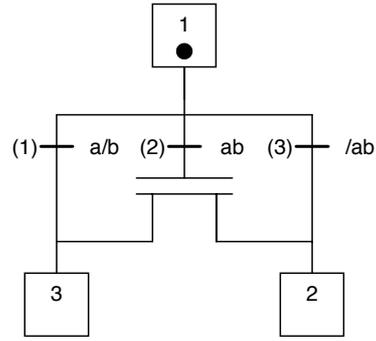

Fig. 7: Eliminating a conflict in a Grafcet

Three different situations can occur: **a**=1 and **b**=0, **a**=0 and **b**=1, **a**=**b**=1. In this case we propose that the grafcet of Fig.5 should be replaced by another equivalent grafcet (Fig. 7) that doesn't contain conflicts, since the receptivity of transitions (1), (2) and (3) are mutually exclusive.

### B. Implementation

The generation of C code from grafcet is not direct. The most important distinction is that grafcet has parallelism (two transitions can be fired simultaneously), which in C must be simulated with several sequential actions.

The method chosen to generate C code uses a generic algorithm, presented in Fig. 8 in Grafcet notation, which will execute any specific grafcet.

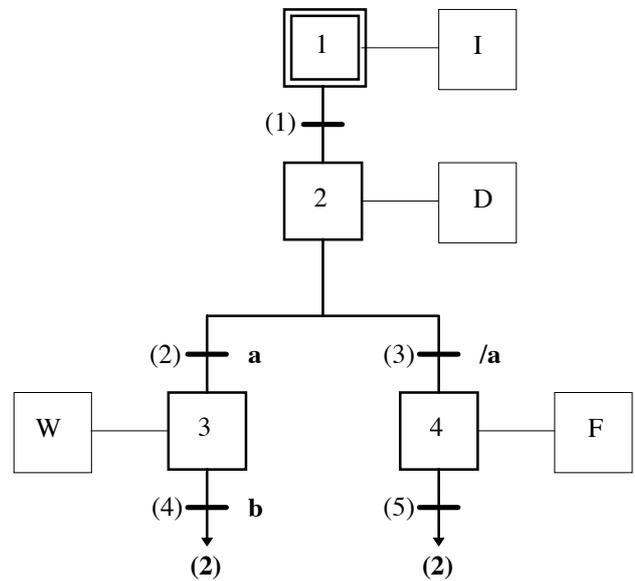

Legend:
- I- Initialisation: activate the initial steps
- D- Determine the set X of fireable transitions
- W- Wait until an external event occur
- F- Fire all fireable transitions
- a- X = { }
- b- occurrence of an external event

Fig. 8. Algorithm for executing a grafcet

Action I initialises the grafcet: activates all initial steps and executes its associated actions. Considering that steps have the attribute *A* which determines the state of the step (active or inactive), activating a step is just setting *A* to true. And since all the associated actions of a step are translated to a C function, executing a step is just invoking that function.

Action D determines the set X of fireable transitions. A transition belongs to this set if all the steps preceding that transition are active and its receptivity is true. This verification is repeated to all transitions of the grafcet. In practice the set X is not created, we simply add a boolean attribute *f* to the transition which is true when that transition is fireable.

The action F - fire all transitions - raises the problem of determine which transition should be fired first. One way to solve this problem is to associate with each step two additional attributes *P* and *T*, where *P* is the priority of that particular step and *T* is the time elapsed since its last activation. The algorithm that schedules the set X of fireable transitions is based on the attributes *P* and *T* of the steps connected after each transition of X.

The action W - wait until an external event occurs - verifies if the user prints a specific character.

## IV. EXTENSIONS

A strategy to specify complex systems is to apply a top-down methodology to the analysis of the system. In the first level, it is defined the overall functionality of the system, aggregating several parts and abstracting specific details. At the succeeding levels, each part of the system is specified separately and with higher detail. At the bottom level, there will be several small sub-systems with a very detailed description. Those sub-systems can not be further divided. This methodology can also be applied to the Grafcet specification of control systems if we extend Grafcet formalism with **macrosteps**. So, in this approach, each step can be a grafcet, which means that a step can represent a sub-system.

A macrostep must satisfy some rules:

(a) it must have only one input step and one output step;
(b) when a macrostep is activated immediately activates is input step;
(c) the output step participates in the enabling of the transitions connected after the macrostep.

A macrostep can be replaced by the grafcet it represents, which is known by macrostep extension. The result of replacing all macrosteps by its extensions is a linear grafcet that represents the complete system.

### A. Implementation

In this top-down methodology, the specification of the control systems defines a hierarchy (tree) of grafcets. In the root of the hierarchy is the most general grafcet, the only one that represents the complete control system. In the implementation, this hierarchy is replaced by a linear structure. In this way, the specification of the system is a set of grafcets: the set of all grafcets presented in the hierarchy. A grafcet that belongs to this set is active if it has active steps or any of its descendent grafcets are active.

So, this set can be divided in two disjoin sets: the set of active grafcets and the set of inactive grafcets.

The execution algorithm of a set of grafcets has the same structure of algorithm presented in Fig. 8, but each particular action has to be extended to treat a set of grafcets. For example, the extended action of action W, which operates on a set of grafcets, is represented by W*.

In the initialisation - action I* - only the root[1] grafcet is active and all of its initial steps are activated. If any of those initial steps is a macrostep then the associated grafcet will be activated.

An important fact is that a macrostep must have three states (active, inactive and complete) to maintain the coherence of the execution algorithm. The additional state *complete* guarantees that, although the macrostep is active, its execution is not yet finished. A macrostep reaches the complete state when its output step becomes active or complete, depending on the output step being a step or macrostep.

To determine the set of fireable transitions - action D* - of the global system it is necessary to verify all the transitions of each one of the active grafcet. As it was mentioned, a transition is fireable if all the preceding steps are active and its receptivity is true.

The action F* - fire all fireable transitions - has two levels of scheduling:

(a) in the first level it chooses one the active grafcets;
(b) in the second level it chooses the transition to be fired of the grafcet selected in the previous item.

There are several ways to choose the next grafcet to be analysed: select the first active grafcet in the sequence, select the grafcet with the lower level[2], add to each grafcet a priority as it was done with the steps in section III.

All of this properties can be used individually or simultaneously to schedule the active grafcets. To choose the next transition that will be fired the method referred in the definition of action in section III is used.

The actual firing of a transition follows the usual rule: all the steps connected after those transitions are activated. When one of this step is an output step of a grafcet $g_i$ then some extra rules must be added:

(i) the grafcet $g_i$ will be deactivated;
(ii) the macrostep associated with $g_i$ becomes complete.

In practice, this means that a sub-system finished its execution.

### B. Interruptions

The fact that each step is formed by several actions indicates the possibility to interrupt its execution at the end of any of its actions. But, generally each action is a unit too small to attend this purpose. It will be more efficient to group related actions, of a particular step, in one unit referred here as task. Considering that each step has several tasks, each of them indivisible, then it will be possible to interrupt any step at the end of the task execution.

---

[1] Notice that the root grafcet will always be active.
[2] The root grafcet has level one.

With this approach, it is possible to control the execution time of a step. Each step will have a maximum number of clock cycles to execute all of its tasks, when this limit is reached the execution of the step is halted and another step will start executing. The disadvantage is that it must be consider the access to variables shared by several steps. Otherwise, when a step that was temporarily halted restarts its execution, it may find variables changed by other steps. So, the great disadvantage of this method is the overhead in

Another approach is to consider that each step is an indivisible unit, this way the execution of its actions can not be interrupted. This solution is easier to manager, but reduces the versatility and applicability of the specification methodology.

*C. Grafcet Libraries*

Until now, it was not considered the possibility of the existence of repeated grafcets in the leaves of the hierarchy. At a first glance, it seams more efficient to maintain a single copy of each different grafcet. But this would limit the number of active identical grafcets to one. The limitation exists because the repeated grafcets are identical, not equal. Each one of the identical grafcets has a different context, the variables can have different values. To allow the existence of several identical grafcets active simultaneously is necessary to maintain a copy for each one of the grafcets.

Almost every programming language offers a set of libraries, that include some executable functions that are commonly used. Within the same perspective it is possible to construct a library of grafcets, that can be reused in various specifications. The reutilization is useful for several reasons. First, its easy to pick a grafcet in the library then develop the same grafcet. Second, any grafcet of the library should be formally verified. That way the grafcets would have a "certification of quality": deadlock free, safe, etc. When the grafcet needed does not belong to library the solution is to develop it. Although, it may exist a similar grafcet in the library that can be worthwhile to adjust to the needed specification.

## V. CONCLUSIONS

The present paper presented a framework for the automatic development of mixed hardware-software controllers, mainly intended for microcontroller based systems.

Presently, our system can generate PLD's and C code. Apart from the traditional ANSI C code, we exploited the use of interrupts and time constrains in this controller's development.

The team involved in this project is also interested in the development of distributed controller with different technology systems: PLC's, microcomputers and microcontrollers. As common strategy, we intend to use a general-purpose specification tool and automatically translate the controller algorithm to include at each controller.

## VI. ACKNOWLEDGEMENTS

Present project is supported by JNICT (Junta Nacional de Investigação Científica) under programme PBICT.